\documentclass[sn-aps,twocolumn,Numbered]{sn-jnl}

\usepackage{graphicx}%
\usepackage{amsmath,amssymb,amsfonts}%
\usepackage{amsthm}%
\usepackage{mathrsfs}%
\usepackage[title]{appendix}%
\usepackage{xcolor}%
\usepackage{textcomp}%
\usepackage{manyfoot}%
\usepackage{booktabs}%
\usepackage{algorithm}%
\usepackage{algorithmicx}%
\usepackage{algpseudocode}%
\usepackage{listings}%

\theoremstyle{thmstyleone}%
\theoremstyle{thmstyletwo}%

\theoremstyle{thmstylethree}%

\begin{document}

\title[Article Title]{Effective exponents near bicritical points}


\author[1,2]{\fnm{Andrey} \sur{Kudlis}}\email{andrewkudlis@gmail.com}

\author*[3]{\fnm{Amnon} \sur{Aharony}}\email{aaharonyaa@gmail.com}

\author[3]{\fnm{Ora} \sur{Entin-Wohlman}}\email{orawohlman@gmail.com}


\author{\fnm{Andrey} \sur{Kudlis$^{\textup{1,2}}$}}\email{andrewkudlis@gmail.com}
\affil[1]{\orgname{Abrikosov Center for Theoretical Physics, MIPT}, \orgaddress{\street{Institutsky lane, 9}, \city{Dolgoprudny}, \postcode{141701}, \state{Moscow Region}, \country{Russia}}}
\affil[2]{\orgname{ITMO University}, \orgaddress{\street{Kronverkskiy prospekt 49}, \city{Saint Petersburg}, \postcode{197101}, \state{State}, \country{Russia}}}

\affil*[3]{\orgdiv{School of Physics and Astronomy}, \orgname{Tel Aviv University}, \orgaddress{\street{Ramat Aviv}, \city{Tel Aviv}, \postcode{6997801}, \state{} \country{Israel}}}




\abstract{The phase diagram of a system with two order parameters, with ${\it n_1}$ and $n_2$ components, respectively, contains two phases, in which these order parameters are non-zero. Experimentally and numerically, these phases are often separated by a first-order ``flop" line, which ends at a bicritical point. For $n=n_1+n_2=3$ and $d=3$ dimensions (relevant e.g. to the uniaxial antiferromagnet in a uniform magnetic field), this bicritical point is found to exhibit a crossover from the isotropic $n$-component universal critical behavior to a fluctuation-driven first-order transition, asymptotically turning into a triple point. Using a novel expansion of the renormalization group recursion relations near the isotropic fixed point, combined with a resummation of the sixth-order diagrammatic expansions of the coefficients in this expansion, we show that the above crossover is slow, explaining the apparently observed second-order transition. However, the effective critical exponents near that transition, which are calculated here, vary strongly as the triple point is approached.

}




\maketitle

\clearpage
\newpage

\section{Introduction}\label{sec1}

  Phase diagrams involving two competing order parameters, with $n^{}_1$ and $n^{}_2$ components, arise in a variety of physical systems. As the temperature is lowered from the disordered phase, one can approach each of these ordered phases, via a phase boundary. These two boundaries meet at a multicritical point, which can be bicritical, tetracritical or triple (Fig. 1).  \cite{KNF,bruce,mukamel} As seen in Fig. 1, the triple point is not really `critical' or `multicritical';  however, it is the meeting point of three first order lines, and it neighbors three phases. It is also reached from the bicritical point. Therefore, we list it together with the `true' multicritical points. The nature of this multicritical point has been under debate for a long time. in particular, the $\epsilon-$expansions for the renormalization group (RG) fixed points (FPs) corresponding to these multicritical points, and the critical exponents  in their close vicinity, have been expanded to fifth order in $\epsilon$ and then resummed for estimating their values in three dimensions ($d=3$).~\cite{vicari}  The resulting numbers agreed with those found by other methods, e.g.,   a resummation of the sixth-order perturbative (divergent) expansions in the original field-theory coefficients at $d=3$~\cite{6loops},  recent bootstrap calculations~\cite{boot}, Monte Carlo simulations~\cite{hasen} and high-temperature series~\cite{butera}.

  All the above calculations derived only the critical exponents {\it at} the FPs. It turns out that in some cases the RG flow away from the unstable isotropic FP (see below) is slow, and sometimes it reaches regions far from this FP, with effective critical exponents which are distinct from those calculated {\it at} that FP. Although such effective exponents were calculated to second order in the coupling constants for the tetracritical point,~\cite{folk} they were not yet calculated for the crossover from bicritical to triple points.
  In Refs. \cite{AEK,AEK2} and \cite{AE}, we extended the above calculations, and presented an accurate RG analysis of such systems close to their multicritical points, confirming that for $n=n_1+n_2=3$ in $d=3$ dimensions it can either be tetracritical, described by the biconical FP, or a triple point, characterized by a slow RG flow away from the isotropic $n-$component FP.
  Since the (stable) biconical FP is very close to the (unstable) isotropic FP, the effective exponents do not change much during the RG flow between them, and the tetracritical point is more or less understood. However, the triple point case requires following the RG trajectory away from the isotropic FP for many RG iterations.  Since this flow is very slow, Refs. \cite{AEK,AEK2,AE} used resummed sixth order expansions of the  derivatives of the $\bar{\beta}$ functions (see below) at the isotropic FP, yielding accurate estimates for the effective exponents, which vary along the RG trajectory. It turned out that the effective exponents change significantly before the triple point is reached. The slow flow indicates that in many cases the triple point may never be reached, and then one may mistakenly identify the multicritical point as a bicritical one, alas with unusual effective exponents.

  As we discuss below, a general four-spin Hamiltonian can be isotropic in spin space, ad it can become anisotropic in many ways. In the description above, the isotropy was broken by the terms which separate the two order parameters, with $n^{}_1$ and $n^{}_2$ components. An alternative symmetry breaking is to add a term with cubic symmetry in spin space, $v\sum_{i=1}^nS_i^4$.~\cite{AAc} As we discuss below, both anisotropies share the same stability exponent.  In Ref. \cite{AEK} we demonstrated the above RG calculation for the cubic case, and showed that all the RG trajectories approach the same universal line. Some effective critical exponents were then calculated for that case in Ref. \cite{AEK2}.
  Reference \cite{AE} used group theoretical arguments, to show that the universal flow line mentioned above is in fact also reached for the case of $O(1)\bigoplus O(2)$ symmetry, relevant to the multicritical points of interest here. The present paper extends those calculations, by calculating the effective exponents for this case.  Qualitatively, the results are similar to those of the cubic case. However, unlike that case, for $2+1$ components we need two sets of exponents, $\nu^{}_{\|,\perp}$ and $\eta^{}_{\|,~\perp}$.

\subsection{The XXZ case}

\begin{figure*}[htb]
\vspace{-1.6cm}
\includegraphics[width=.96\textwidth]{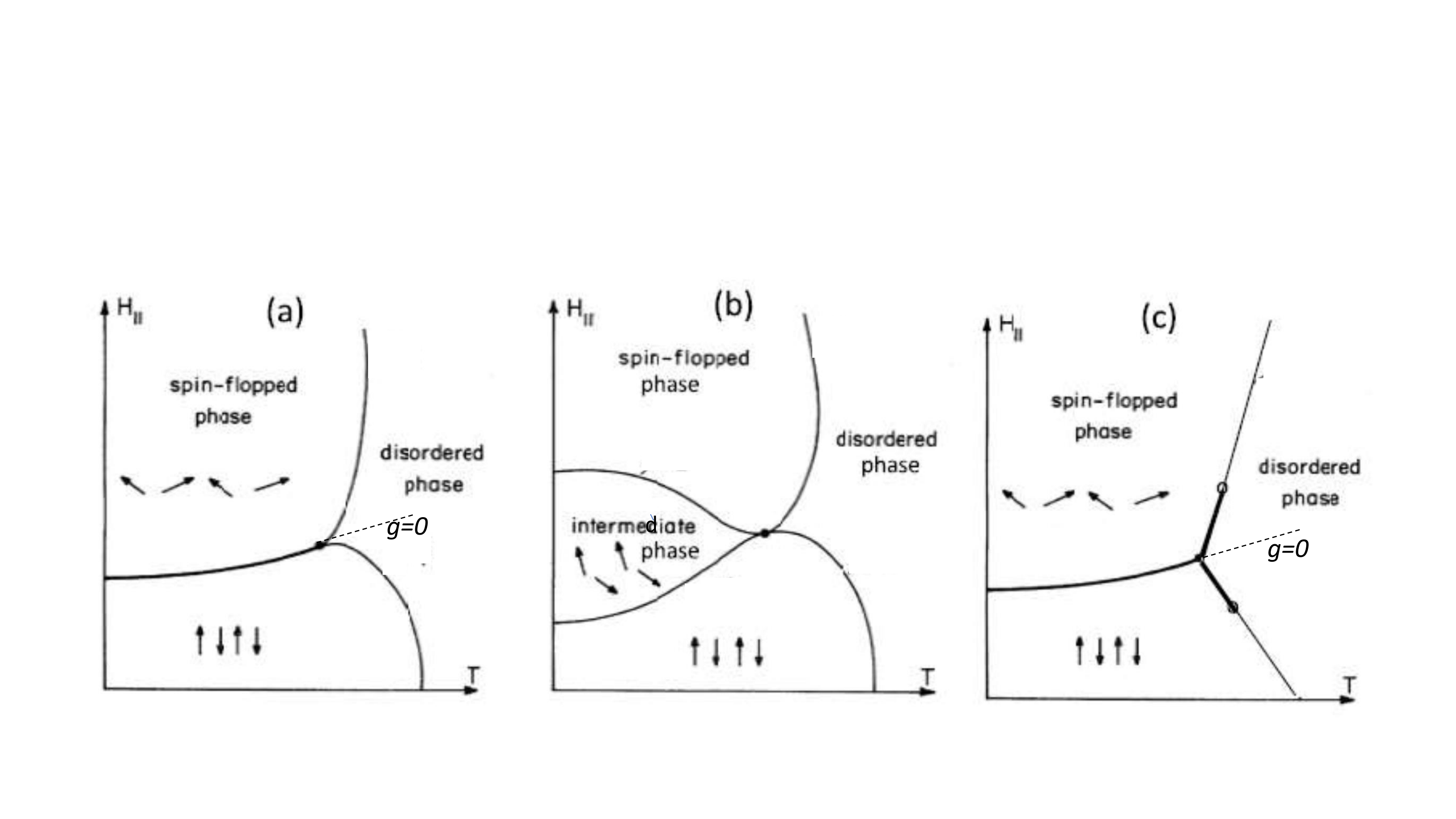}
\vspace{-1cm}
\caption{  Possible phase-diagrams for the XXZ antiferromagnet in a longitudinal magnetic field. (a) Bicritical phase diagram.  (b) Tetracritical phase diagram. (c) Diagram with a triple point.  Thick lines - first-order transitions. Thin lines - second-order transitions. The first-order transition lines between the ordered phases and the disordered paramagnetic phase turn into second order ones  at tricritical points (small empty circles). After Refs. \cite{bruce,mukamel}. }
\label{1}
\end{figure*}

Most of our results refer to the most ubiquitous example of the uniaxially anisotropic XXZ antiferromagnet in a uniform magnetic field, with $n_1=2,~n_2=1$. Other examples are mentioned in Refs. \cite{AEK,AE}.
A uniaxially anisotropic XXZ antiferromagnet has long-range  order (staggered magnetization)  along its easy axis, Z. A magnetic field $H^{}_\parallel$ along that axis causes a spin-flop transition into a phase with order in the transverse plane, plus a small ferromagnetic order along Z. Experiments~\cite{king,Shapira} and Monte Carlo simulations on three-dimensional lattices ~\cite{selke0,selke,landau} typically find an apparent {\it bicritical} phase diagram in the temperature-field $T-H^{}_\parallel$ plane [Fig. \ref{1}(a)]: a first-order transition line between the two ordered phases, and two second-order lines between these phases and the disordered (paramagnetic) phase, all meeting at an apparent {\it bicritical point}.

The field theoretical analysis is based on the Ginzburg-Landau-Wilson (GLW) Hamiltonian density~\cite{KNF}, ${\cal H}({\bf r})={\cal H}^{}_0+U^{}_2+U^{}_4$, with
\begin{equation}
{\cal H}^{}_0({\bf r})=\big(\mid{\boldmath{\nabla}}{\bf S}\mid^2+r\mid{\bf S}\mid^2\big)/2,
\end{equation}
\begin{equation}
U^{}_2=g\big[ \mid S^{}_\|\mid^2-\mid{\bf S}\mid^2/3\big],
\end{equation}
\begin{align}
U^{}_4=&u^{}_\|\mid S^{}_\|\mid^4+ u^{}_\perp \mid{\bf S}^{}_\perp\mid^4\nonumber\\
&+2 u^{}_\times \mid S^{}_\|\mid^2\mid{\bf S}^{}_\perp\mid^2,
\label{U4}
\end{align}
with the local three-component ($n=3$) staggered magnetization,  ${\bf S}({\bf r})\equiv \big({\bf S}^{}_\perp({\bf r}),S^{}_\|({\bf r})\big)$.
Here, $r=(T-T^{}_c)/T^{}_c$ measures the distance from the critical temperature. For $g=0$ and $u^{}_\|=u^{}_\perp=u^{}_\times=u$, ${\cal H}$ reduces to the isotropic Wilson-Fisher Hamiltonian~\cite{wilson,RG,DG}, which has an (isotropic) FP at $u=u^{\ast,I}$.

\subsection{Renormalization group}

After $\ell$ iterations of the RG, the size of the system becomes $L\Rightarrow L(\ell)=L/e^\ell$, and the correlation length becomes $\xi\Rightarrow \xi(\ell)=\xi/e^\ell$ (in units of the lattice constant). In addition, various scaling operators, ${\cal O}^{}_a$, e.g., $S^{}_i$ and $S^2_i$, are rescaled, ${\cal O}^{}_a=Z^{}_a(\ell){\cal O}^{}_a$, with scaling factors $Z^{}_a$ which depend on $\ell$.~\cite{wilson,RG,DG} This maps ${\cal H}$ to ${\cal H}(\ell)$, with renormalized coefficients, e.g., $r(\ell),~g(\ell),~u^{}_i(\ell)$. The RG studies the recursion relations which yield the trajectories of these coefficients in the parameter space, and the fixed points of these trajectories.

Using $\xi\sim \mid t\mid^{-\nu}$, where $t$ is the scaling field related to $r$ and $\nu$ is the `isotropic' exponent for the correlation length, we continue iterating until $\min[L(\ell^{}_f),~\xi(\ell^{}_f)]=1$. Thus, $\ell^{}_f$ is the smaller of $\ln L$ and $-\nu \ln\mid t\mid$. In real or numerical simulations, $\ell^{}_f$ grows with the system's finite size and/or by the temperature range which is used.
At $\ell=\ell^{}_f$, all fluctuations have been eliminated and one can solve the problem using the mean-field Landau theory~\cite{RG}.
Reaching $\ell^{}_f$ requires the {\it full RG flow} of the system's Hamiltonian.  

The main RG calculations involve the recursion relations for the
three coefficients in $U^{}_4$, 
\begin{align}
\frac{\partial u^{}_i}{\partial \ell}=\bar{\beta}_i[\epsilon,~u^{}_{\|},~u^{}_\perp,~u^{}_\times].
\label{barbeta}
\end{align}
When $\partial u^{}_i/\partial\ell=0$, these recursion relations yield four FPs, the Gaussian ($u^{}_{\|}=u^{}_\perp=u^{}_\times=0$, the isotropic ($u^{}_{\|}=u^{}_\perp=u^{}_\times=u^{\ast,I}$), the decoupled ($u^{}_\times=0$) and the biconical FPs.
In three dimensions, the locations of these fixed points, and the critical exponents in their vicinity,  took some time to find, mainly because of difficulties to extrapolate the recursion relations down from $d=4$ to $d=3$, given as divergent series in $\epsilon=4-d$  and in the $u$'s. One way to overcome this is to use resummation techniques, e.g.,  by taking into account the singularities of the series' Borel transforms~\cite{vicari},  and extrapolating the results to $\epsilon=1$. These showed that for $n=d=3$
the only stable FP is the biconical one, which is very close to the unstable isotropic one. The results for the critical exponents agreed with  a resummation of the sixth-order perturbative (divergent) expansions in the original field-theory coefficients at $d=3$~\cite{6loops}, with recent bootstrap calculations~\cite{boot}, with Monte Carlo simulations~\cite{hasen} and with high-temperature series~\cite{butera}.

Since the biconical and isotropic FPs are very close to each other, we decided to study in detail the recursion relations in the vicinity of the isotropic FP. To first order in the $u$'s, we need the three exponents of its stability in the space of the $u$'s. It turns out that the associated eigenvectors of these recursion relations are given by group theory.  In our case, [$O(n)\Rightarrow O(n^{}_1)\bigoplus O(n^{}_2)$],
 group theory shows that these eigen-operators  are~\cite{wegner,zan,vicari,hasen,vicrev}
\begin{align}
&{\cal P}^{}_{4,0}\equiv\mid{\bf S}\mid^4,\   {\cal P}^{}_{4,2}\equiv \mid{\bf S}\mid^4[x-n^{}_1/n],\nonumber\\
&{\cal P}^{}_{4,4}\equiv\mid{\bf S}\mid^4\big[\frac{n^{}_1n^{}_2}{(n+2)(n+4)}\nonumber\\
&+x(1-x)
-\frac{n^{}_1(1-x)+n^{}_2 x}{n+4}
\big],\label{P4}
\end{align}
where $x=S^{2}_{\|}/\mid{\bf S}\mid^2$.
For $n=3=2+1$, the corresponding stability exponents (agreed by all the extrapolations) are
 ~\cite{boot}
 \begin{align}
 \lambda^{}_0\approx -0.78,\ \lambda^{}_2\approx -0.55,\ \lambda^{}_4\approx 0.01.
 \label{exps}
 \end{align}

Rewriting Eq. (\ref{U4}) as
\begin{align}
U^{}_4&=(u^{\ast,I}+p^{}_0) {\cal P}^{}_{4,0}+p^{}_2{\cal P}^{}_{4,2}\nonumber\\
&-p^{}_4{\cal P}^{}_{4,4},
\label{U4n}
\end{align}
the linear recursion relations for the coefficients $p^{}_i$ and their solutions are
\begin{align}
 dp^{}_i/d\ell&\approx\lambda^{}_ip^{}_i\ \ \Rightarrow\ \ p^{}_i(\ell)=p^{}_i(0)e^{\lambda^{}_i\ell},\nonumber\\
 &i=0,~2,~4.
 \label{pil}
 \end{align}
Group theory also identifies the eigenvectors, which yield the exact relations
\begin{align}
&\delta u^{}_{\|}=p^{}_0 + (70 p^{}_2 + 24 p^{}_4)/105,\nonumber\\
&\delta u^{}_\perp= p^{}_0 - (35 p^{}_2 - 9 p^{}_4)/105,\nonumber\\
&\delta u^{}_\times=p^{}_0 + (35 p^{}_2 - 72 p^{}_4)/210,
\label{uuu}
\end{align}
where  $\delta u^{}_i=u^{}_i-u^{\ast,I},~i=\|,~\perp,~\times$.
These linear expressions can easily be generalized to any $n^{}_1$ and $n^{}_2$.

 The largest (negative) exponent $\lambda^{}_0$ corresponds to the stability within the $O(3)-$symmetric case, ${\cal P}^{}_{4,0}$. In our case, the exponent $\lambda^{}_2$ corresponds to a term which splits the $O(3)$ isotropic symmetry group  into $O(1)\bigoplus O(2)$.
Similar to $U^{}_2$, ${\cal P}^{}_{4,2}$ `prefers' ordering of $S^{}_{\|}$ or of ${\bf S}^{}_\perp$. At the multicritical point the $O(3)$ symmetry must be preserved, and therefore below we set $p^{}_2=0$.

 Since $\lambda^{}_4>0$, the isotropic FP is unstable at $d=3$, and the small $\lambda^{}_4>0$ represents the slow crossover away from the isotropic FP. Examples of the flow trajectories in the $p^{}_0-p^{}_4$ are show in Fig. \ref{fig:flows_up}.  For some range of the parameters, the RG flow reaches slowly the less symmetric FP (biconical or cubic), and the bicritical phase diagram should be replaced by the tetracritical diagram
~\cite{AAco}. Alternatively, the iterations first flow slowly and remain near the isotropic FP, and then flow quickly towards the triple point, beyond which the transition becomes fluctuation driven first order. Neither of these agrees with the experiments or the simulations.

Indeed, the same value of $\lambda^{}_4$ was also found for the crossover from the isotropic to the cubic FP.~\cite{AAc,aaDG,eps6,6loops,boot,hasen,AEK}.
In fact, the most general space of even quartic terms contains fifteen coefficients $u^{}_{abcd}$, which split into groups of $1+5+9$. All the coefficients in each group have the same stability exponent. The cubic $v$ and our $p^{}_4$ both belong the the group of size $9$, and therefore their qualitative behavior is similar.

\section{Our calculation}

Expanding Eq. (\ref{barbeta}) to second order in the deviations from the isotropic FP,
\begin{align}
\frac{\partial u^{}_i}{\partial \ell}=\sum C^{}_{abc}L^i_{abc} (\delta u^{}_{\|})^a(\delta u^{}_{\perp})^b(\delta u^{}_{\times})^c,
\label{RR}
\end{align}
 with the integers $a,b,c=0,1,2,~~1\leq a+b+c\leq2 $, $C^{}_{abc}=(\delta^{}_{a2}+\delta^{}_{b2}+\delta^{}_{c2})/2$, and e.g.
\begin{align}
L^i_{100}=&{\rm Resum}\Big[\frac{\partial \bar{\beta}^{}_i}{\partial u^{}_{\|}}\Big]^{}_{\{u^{}_{j}=u^{\ast,I}\}},\nonumber\\
L^i_{011}=&{\rm Resum}\Big[\frac{\partial^2\bar{\beta}^{}_i}{\partial u^{}_{\perp} \partial u^{}_\times}\Big]^{}_{\{u^{}_{j}=u^{\ast,I}\}},\nonumber\\
L^i_{020}=&{\rm Resum}\Big[\frac{\partial^2\bar{\beta}^{}_i}{\partial (u^{}_{\perp})^2}\Big]^{}_{\{u^{}_{j}=u^{\ast,I}\}}.
\end{align}
The derivatives are calculated at the isotropic FP. Using the known $\epsilon$-expansion of $u^{\ast,I}$, they are obtained as sixth order polynomials in $\epsilon$, which are then resummed - using the methods explained e.g. in Ref. {\cite{AEK}. The resulting values are listed in the Supplementary Material~\cite{SM}.

Diagonalizing the $3\times 3$ matrix of the linear terms in Eq. (\ref{RR}) we recover Eq. (\ref{pil}). The resulting numerical values of the eigenvalues $\lambda^{}_i$'s, are found to be very close to the numbers in Eq. (\ref{exps}), which was based on the full sixth order recursion relations ad on other accurate calculations. Also,  the eigenvectors turn out to be very close to the exact Eq. (\ref{P4}).
We next use Eqs. (\ref{uuu}) to replace Eq. (\ref{RR}) by quadratic equations for the $p$'s. As stated above, at the multicritical point we set $g=0$ and $p^{}_2=0$.
 In this case, we are left with the RG flow in the $p^{}_0-p^{}_4$ plane, and we generalize the calculation of Ref. \cite{AEK}. We first replace $p^{}_0$ by the nonlinear scaling field $q^{}_0$, which obeys
 \begin{align}
 dq^{}_0/d\ell\approx\lambda^{}_0q^{}_0\ \Rightarrow\ q^{}_0(\ell)=q^{}_0(0)e^{\lambda^{}_0\ell}
 \label{qil}
 \end{align}
to all orders in the $p$'s. This is achieved by writing
 \begin{align}
 q^{}_0&=p^{}_0+z^{}_{20}p^2_0+z^{}_{11}p^{}_0p^{}_4\nonumber\\
 &+z^{}_{02}p^2_4+\dots,
 \label{NL}
 \end{align}
and choosing the $z$'s so that the higher order coefficients in $dq^{}_0/d\ell$ vanish.~\cite{SM}
Substituting Eq. (\ref{qil}) into $dp^{}_4/d\ell$, we find
\begin{align}
 \frac{dp^{}_4}{d\ell}=\lambda^{}_4 p^{}_4+A\lambda^{}_0q^{}_0(\ell)p^{}_4-Bp^2_4,
 \label{eqp4}
 \end{align}
 where $B=0.144(11)$ and  $A=2.369(67)$.~\cite{SM}
 Rewriting this as a differential equation in $1/p^{}_4$,
writing $x=e^{\lambda^{}_0 \ell}$ and
$1/p^{}_4(x)=e^{-A x} x^{-\lambda^{}_4/\lambda^{}_0}W(x)$,
yields
\begin{align}
dW/dx=[B/\lambda^{}_0]e^{\tilde{A}x}x^{\lambda^{}_4/\lambda^{}_0-1},
\end{align}
with the solution
\begin{align}
&W(\ell)=\frac{e^{\tilde{A}}}{p^{}_4(0)}+
\frac{(-\tilde{A})^{-\lambda^{}_4/\lambda^{}_0}B}{\lambda^{}_0}\times\nonumber\\
&
\Big(\Gamma[\lambda^{}_4/\lambda^{}_0,-\tilde{A}]
-\Gamma[\lambda^{}_4/\lambda^{}_0,-\tilde{A}e^{\lambda^{}_0\ell}]\Big),
\label{solw}
\end{align}
where $\tilde{A}=Aq^{}_0(0)$, and $\Gamma[s,z]$   
is the incomplete gamma function. 
From Eq. (\ref{NL}), the quadratic approximation gives
\begin{align}
    p_0\approx q_0 - z_{20} q_0^2-z^{}_{11}q^{}_0p^{}_4 - z_{02} p^{2}_4 . \label{eqn:p0_beh}
\end{align}
This equation, together with Eqs. (\ref{qil}) and (\ref{solw}), generate the trajectories in the $p^{}_0-p^{}_4$ plane, shown in Fig.~\ref{fig:flows_up}.

\begin{figure}[b]
\centering
\includegraphics[width=1.0\linewidth]{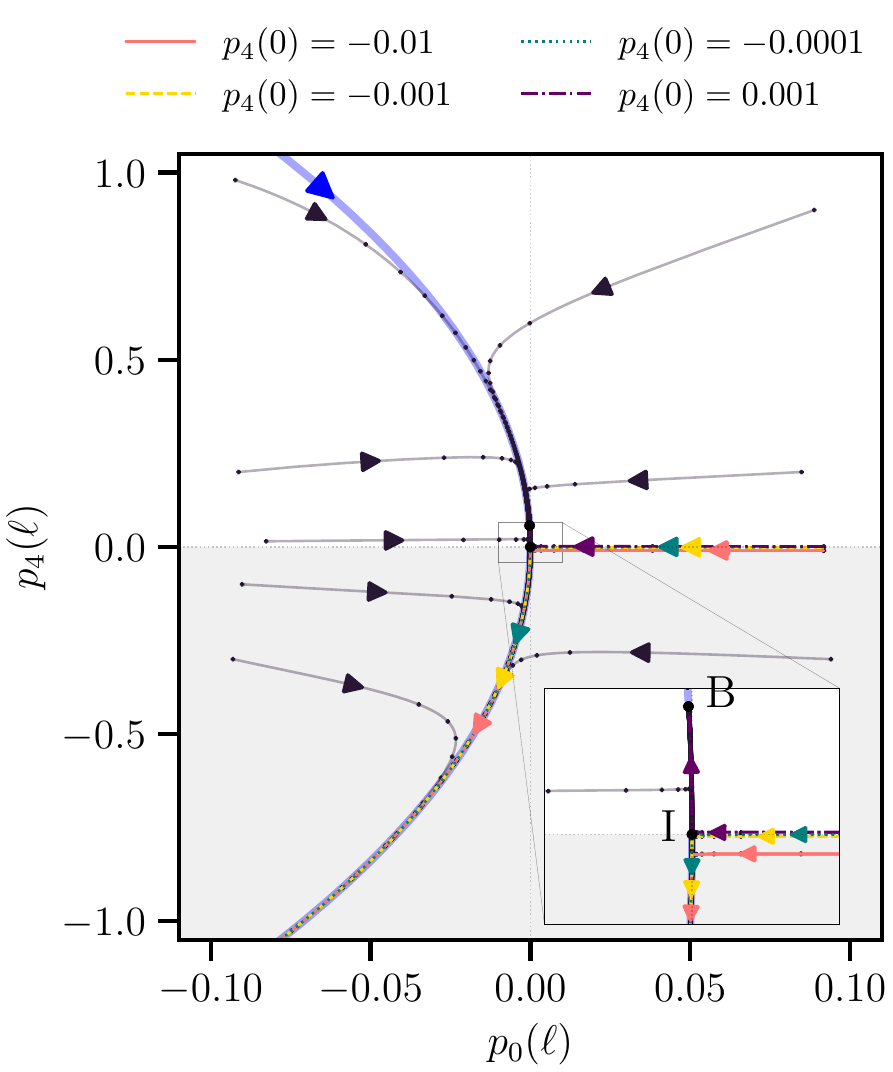}
\caption{(color online) Flow trajectories in the $p^{}_0-p^{}_4$ plane for several initial points. Four of them (presented within the legend bar) are used as initial points for analysis of effective exponents. The trajectories are constructed on the basis of equations~\eqref{eqn:p0_beh} and (\ref{solw}). The dots indicate integer values of $\ell$. The blue thick line is the  universal asymptotic line, when $q^{}_0=0$. The letters $\textup{B}$ and $\textup{I}$ correspond to the biconical and  isotropic ($p^{\ast,I}_4=0$) fixed points, respectively.}
\label{fig:flows_up}
\end{figure}

For large $\ell$, $x=e^{\lambda^{}_0\ell}$
is small, and  $\Gamma[s,z]=-z^s/s+{\cal O}[1]$, so that $\Gamma[\lambda^{}_4/\lambda^{}_0,-Ax]\propto e^{\lambda^{}_4\ell}$.  This result can be obtained directly:   For $\ell>\ell^{}_1$  we can neglect $q^{}_0(\ell)$  in Eq. (\ref{eqp4}). The solution to this equation is then
\begin{align}
p^{}_4(\ell)=\frac{p^{}_4(\ell^{}_1)e^{\lambda^{}_4(\ell-\ell^{}_1)}}{1+B p^{}_4(\ell^{}_1)\big[e^{\lambda^{}_4(\ell-\ell^{}_1)}-1\big]/\lambda^{}_4}.
\label{vvv}
\end{align}
Indeed, this dependence of $p^{}_4$ on $\ell$ is approached for all $p^{}_0(0)$ and large $\ell$. The corresponding asymptotic line, given by $q^{}_0=0$, depends only on the coefficients $z^{}_{ab}$ in Eq. (\ref{NL}) and on $B$. Since these numbers all follow from the derivatives of the $\bar{\beta}$-functions at the isotropic FP, they are all {\it universal}. Therefore the asymptotic line is also universal.

Since $\lambda^{}_4$ is very small, the variation of the second term in the denominator with $\ell$ is slow.
In our case, $B>0$, implying a slow variation in $p^{}_4(\ell)$ for positive $p^{}_4$, approaching the biconical FP $p^{\ast,B}_4=\lambda^{}_4/B=0.057(51)$. This value of $p^{\ast,B}_4$ agrees with the full solution of the original sixth order recursion relations, justifying our quadratic approximation. In contrast, for $p^{}_4(0)<0$, $p^{}_4(\ell)$ becomes more and more negative. At first it decreases slowly, not far from the isotropic FP, but when $e^{\lambda^{}_4 \ell}$ becomes of order unity this decrease becomes faster (the points at integer $\ell$ become less dense),  and $p^{}_4(\ell)$ diverges at $\ell=\ell^{}_2$, when $\big[e^{\lambda^{}_4(\ell^{}_2-\ell^{}_1)}-1\big]/\lambda^{}_4\approx (\ell^{}_2-\ell^{}_1)=-1/[Bp^{}_4(\ell^{}_1)]$. This value is larger for smaller $\mid p^{}_4(\ell^{}_1)\mid$
 [and therefore also for smaller $\mid p^{}_4(0)\mid$].  Within our quadratic approximation, we are not allowed to follow this solution beyond some finite value, say $p^{}_4<-.8$. Although Figs. \ref{fig:flows_up} and \ref{2} show larger values of $\mid p^{}_4(\ell)\mid$, those parts can only be take qualitatively; it is reasonable that a full solution will also continue downwards, on the asymptotic trajectory.

 For large $\ell$, away from criticality ($T\ne T^{}_c$), we can neglect $p^{}_0(\ell)$ and approximate the free energy by its Landau expression,
 \begin{align}
 &F=r(\ell)\mid {\bf S}\mid^2/2+ \mid {\bf S}\mid^4\big(u^{\ast,I}-\nonumber\\
 &p^{}_4(\ell)[2/35+x(1-x)-(1+x)/7]\big).
 \end{align}
 For $p^{}_4(\ell)<0$, the last term in minimal at $x=1$, when the square brackets become $-8/35$. However, the total quartic term becomes negative when $p^{}_4=-35u^{\ast,I}/8\approx -1.75$, and then the transition becomes first order, identifying the triple point. Interestingly, this is exactly the same value found for the cubic case.~\cite{AEK}. 

\begin{figure}[t]
\includegraphics[width=0.9\linewidth]{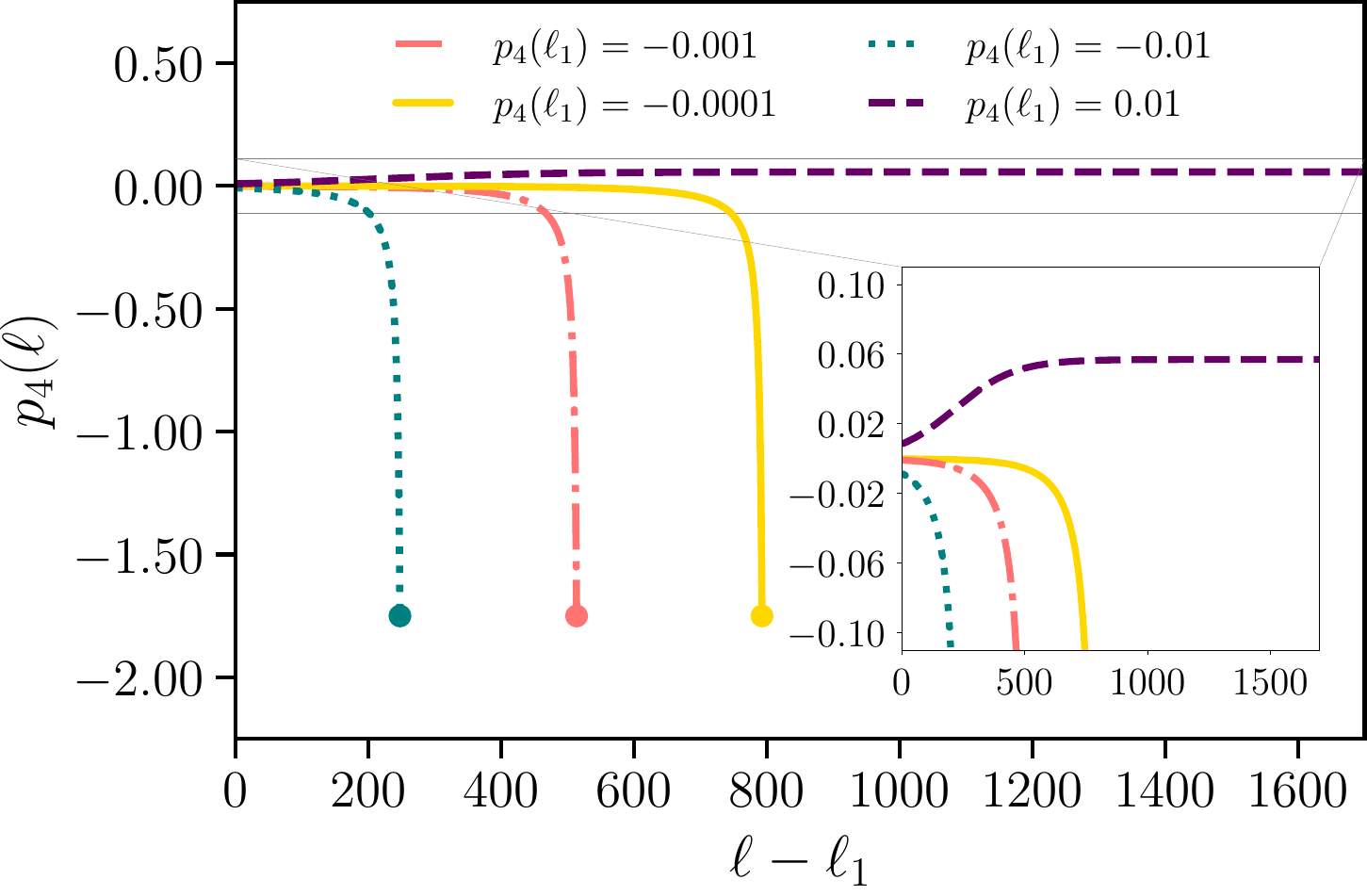}
\caption{(color online)  The function $p^{}_4(\ell)$, Eq. (\ref{vvv}) for $B=0.144$, $\lambda_4=0.0082$, $q^{}_0(\ell^{}_1)=0$ and different $p^{}_4(\ell^{}_1)$, especially highlighted in the flow diagram. Thick dots correspond to the values of $\ell$, when the corresponding $p^{}_4=-35 u^{\ast,I}/8=-1.75$. At this (approximate) point the transition becomes first order and the bicritical point becomes a triple point.}
\label{2}
\end{figure}

\section{Effective Critical Exponents}

So far, we discussed only the recursion relations for the quartic spin terms, $U^{}_4$. To obtain the physically measurable critical exponents we now return to the quadratic terms, which contain $r(\ell)$ and $g(\ell)$. Although these are the correct linear scaling fields, it is also convenient to write the quadratic terms as
\begin{align}
&\big(\mid{\boldmath{\nabla}}{\bf S}\mid^2+r^{}_{\|}\mid S^{}_{\|}\mid^2+r^{}_\perp\mid{\bf S}^{}_\perp\mid^2\big)/2,
\end{align}
where $r^{}_{\|}=r-n^{}_1g$ and $r^{}_\perp=r+n^{}_2 g$ are the temperature parameters associated with the spins $S^{}_{\|}$ and ${\bf S}^{}_\perp$.

At the isotropic FP, $g=0$, and the two $r$'s have the same recursion relation. However, these relations change at order $u^2_i$, ~{\cite{KNF,vicari,folk}
\begin{align}
\partial r^{}_i/\partial\ell=(2-\eta^{}_i)r^{}_i+O[\{u^{}_i\}]
\end{align}
The different prefactors of the first terms result from the different rescaling factors of the operators $S^2_{\|}$ and $\mid{\bf S}^{}_\perp\mid^2$, with $\eta^{}_i=O[u^2_i]$. Technically, these are obtained from keeping the coefficient of the gradient term equal to 1.

\begin{figure}[t!]
    \centering
    \includegraphics[width = .89\linewidth]{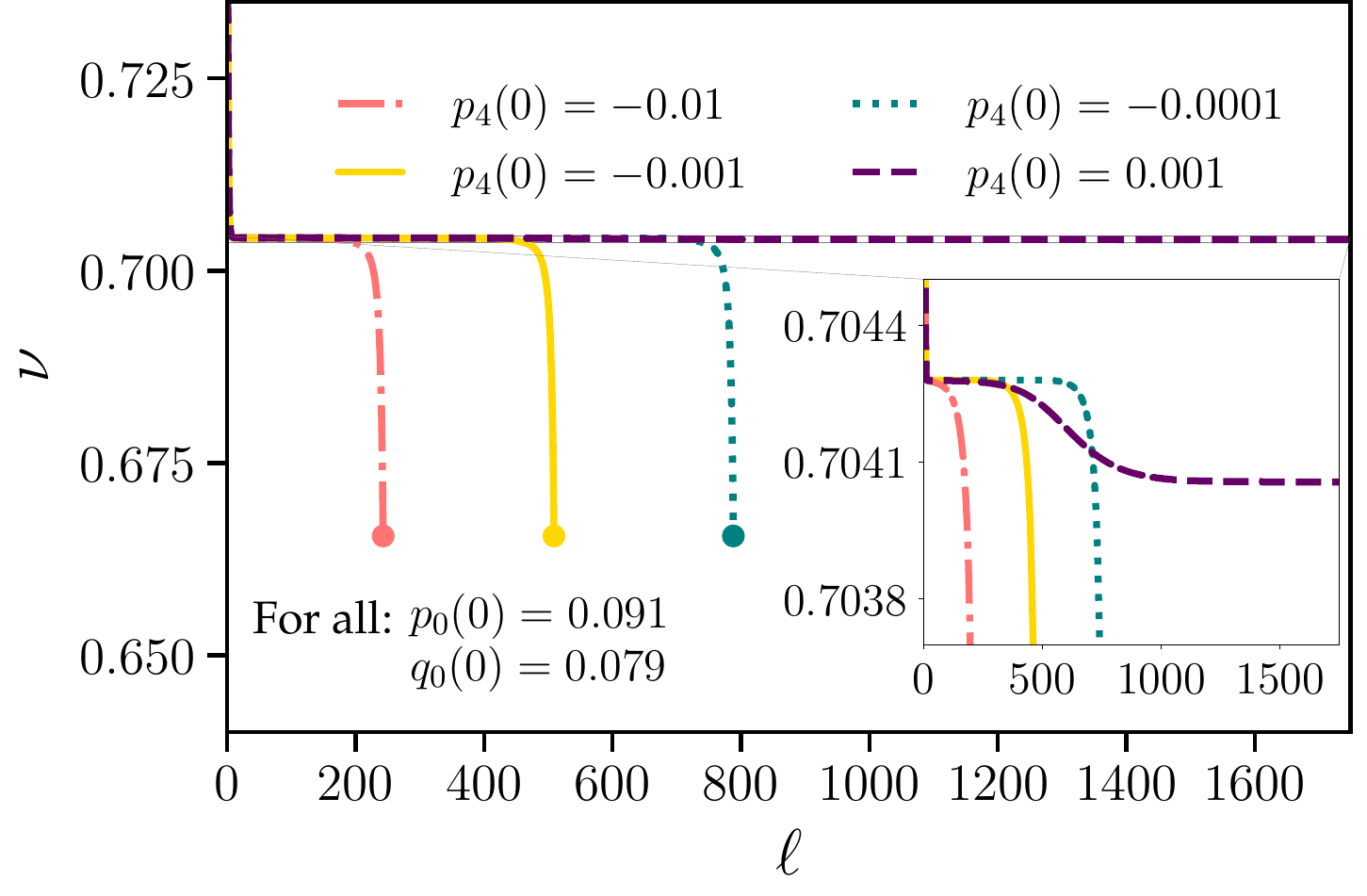}
    \includegraphics[width = .89\linewidth]{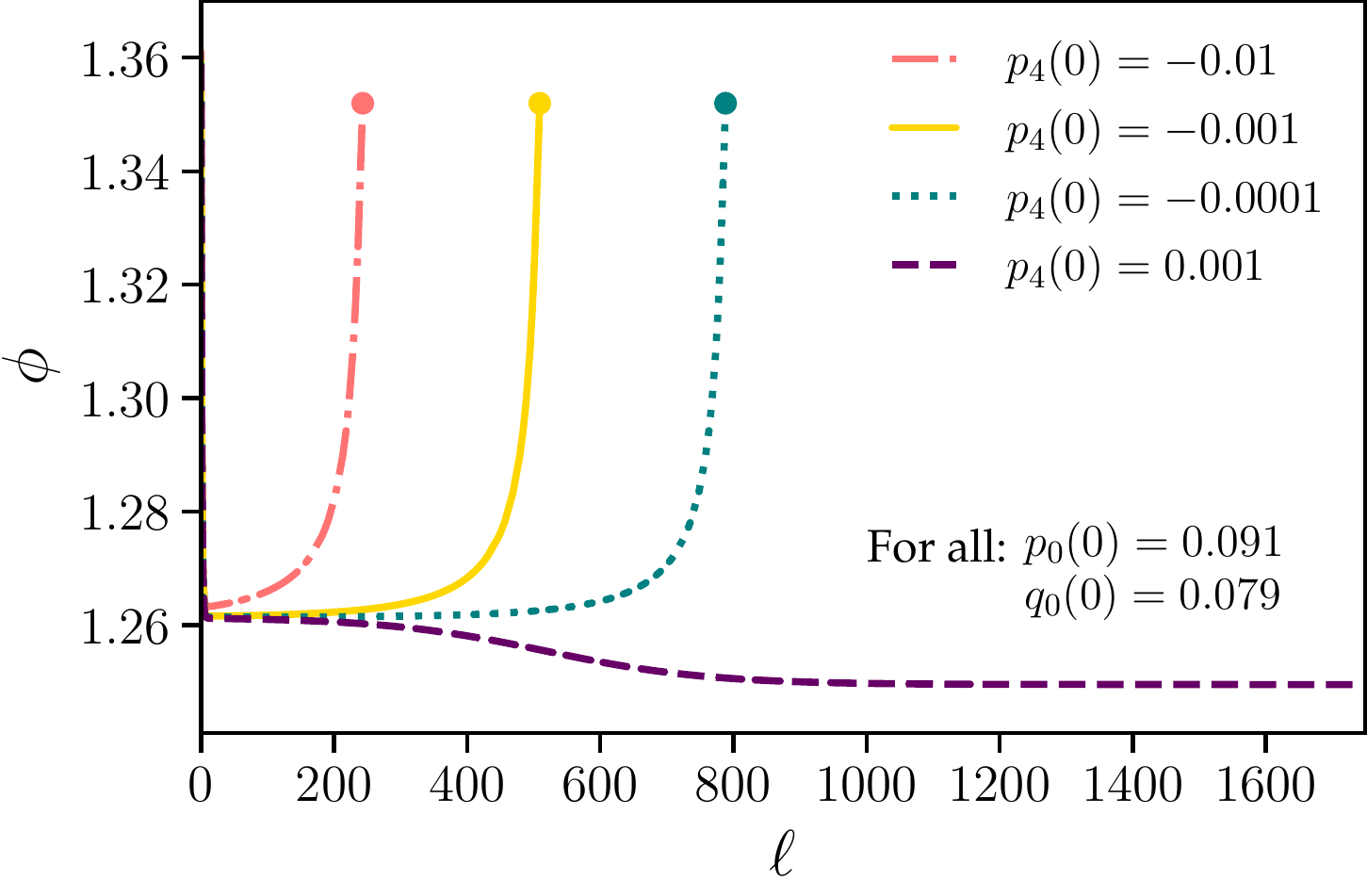}
    \includegraphics[width = .89\linewidth]{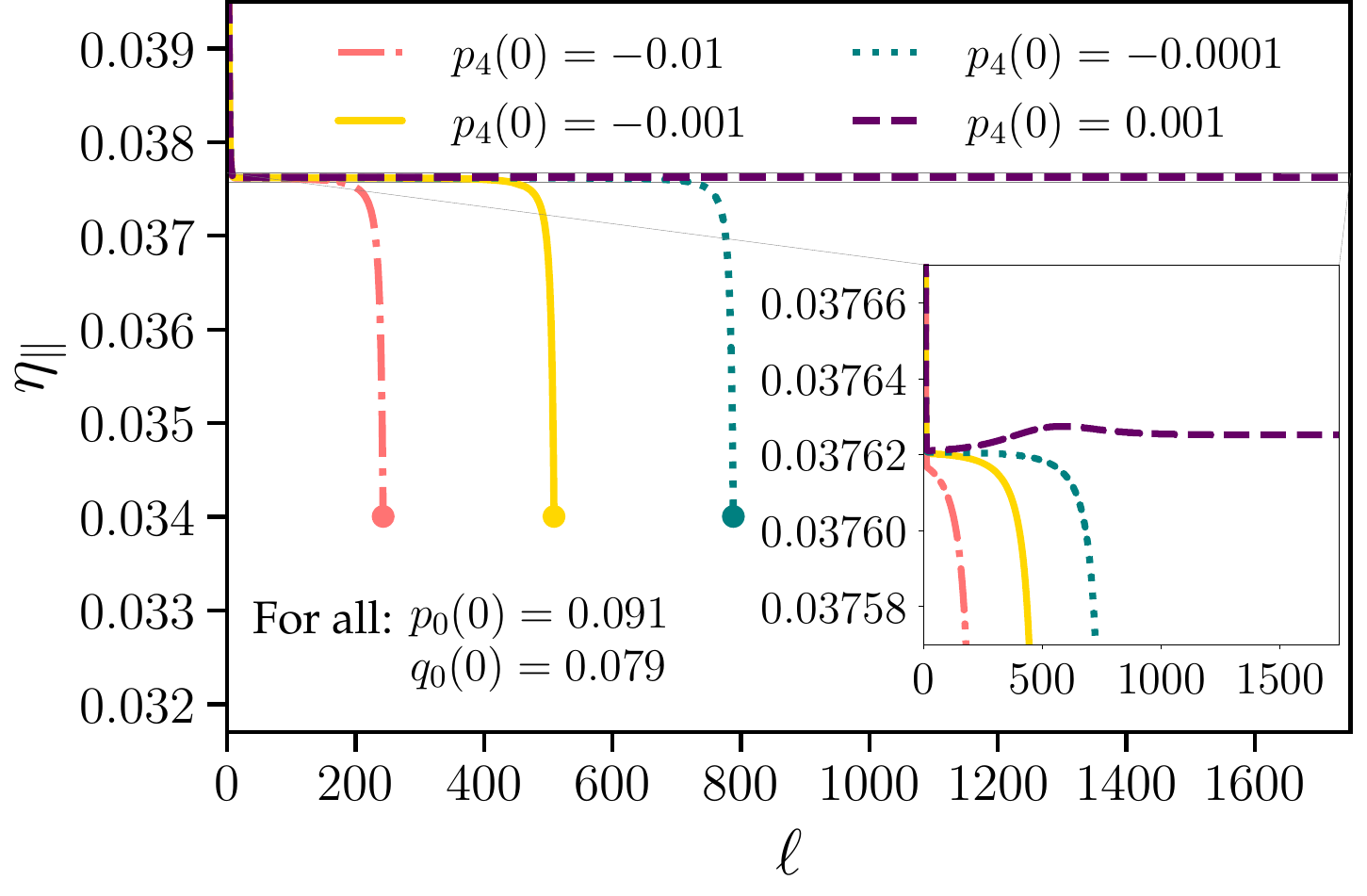}
    \includegraphics[width = .89\linewidth]{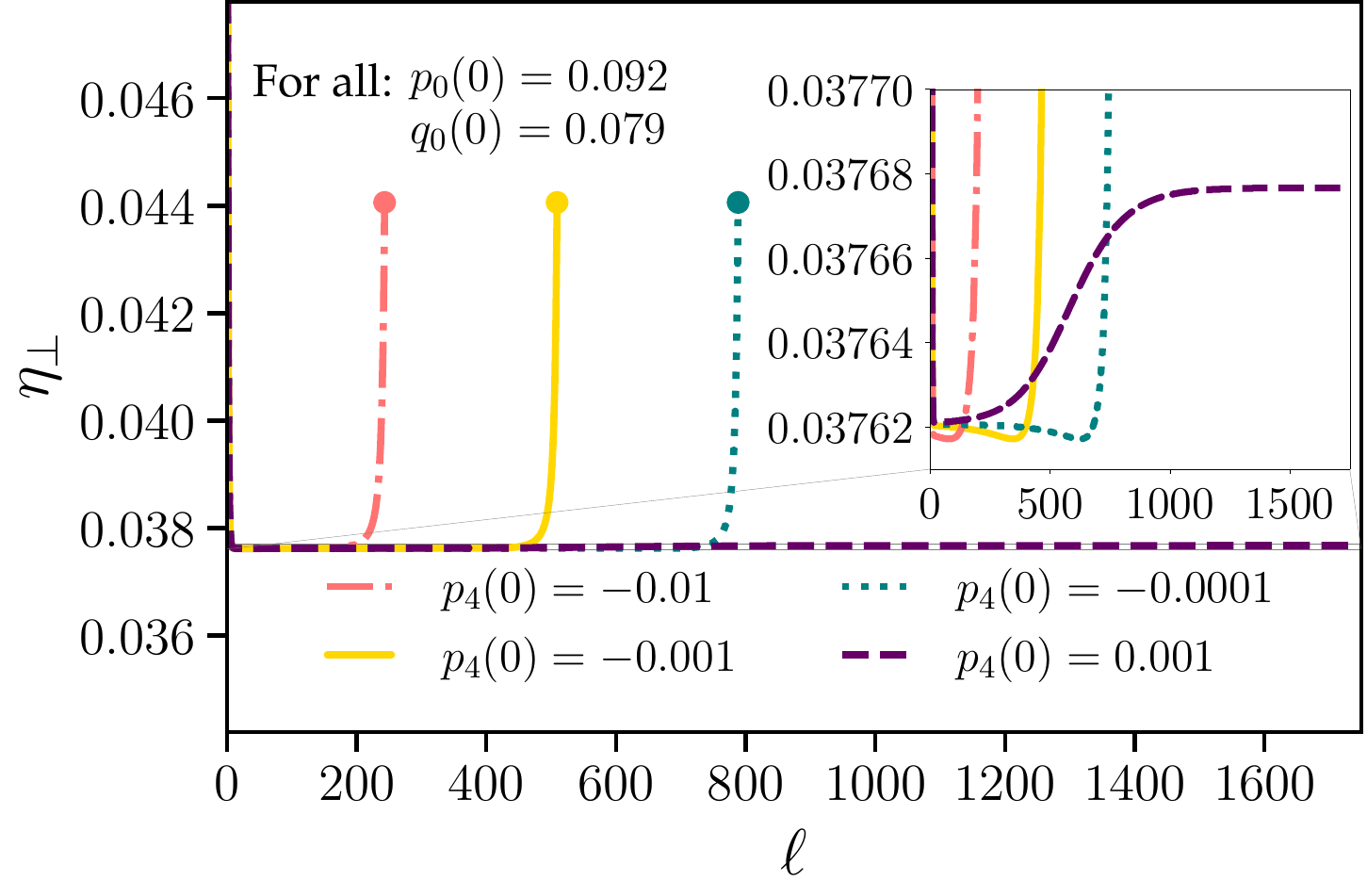}
    \caption{Dependence of effective critical exponents $\nu$, $\phi$, $\eta_{\parallel}$, and $\eta_{\perp}$ on the RG flow parameter $\ell$, based on the four colored trajectories in Fig. \ref{fig:flows_up}. Different lines correspond to different initial values  $p_4(0)$ and $p_0(0)$. 
    Due to our quadratic approximation, all the plots stop at $p_4(\ell)=-.8$.}
    \label{exps}
\end{figure}

The  $\eta^{}_i$'s  describe the power law decay of the correlation functions at $T^{}_c$. Generally,
the Fourier transform of the two point correlation function,   $G^{}_{ii}({\bf q})\equiv\langle S^{}_i({\bf q})q)S^{}_i(-{\bf q})\rangle$, obeys the generalized scaling relation ~\cite{MEFAA}
\begin{align}
G^{}_{ii}(t,{\bf q})&=e^{(2-\eta^{}_i)\ell}G^{}_{ii}(e^{\ell/\nu} t,\nonumber\\
&~e^{\ell \phi/\nu}g,~e^\ell {\bf q}),
\label{GG}
\end{align}
where $i=\|$ or $\perp$. Here, $t$ and $g$ are the nonlinear scaling fields related to $r$ and $g$. Note: the exponent $\nu$ is still isotropic; there is only one correlation length at the multicritical point. All the other critical exponents can be derived from these exponents, e.g., $\gamma^{}_i=(2-\eta^{}_i)\nu$ (obtained from $G^{}_{ii}(t,0)\sim t^{-\gamma^{}_i}$) and $\alpha=2-d\nu$.
The exponent $\phi$ is the crossover exponent connected with the flow away from the multicritical point towards positive or negative $g$ (upper or lower phases in Fig. \ref{1}.
 All the effective exponents $1/\nu,~\phi/\nu,~\eta^{}_{\|}$ and $\eta^{}_\perp$ can be derived from the recursion relations of the two-points correlation functions.~\cite{comvic}
 As before, we expand the corresponding renormalization factors in powers of $\epsilon$ and the $p^{}_i$'s, express the coefficients in terms of $u^{\ast,I}$ and resum these sixth-order $\epsilon-$expansions to obtain expansions of these four exponents to second order in the $p^{}_i$.


Given the solutions $p^{}_i(\ell)$ from the previous section, we have derived the four effective exponents as functions of $\ell$. The effective exponents are given by
\begin{align}
\chi^{}_{i}=&\chi^{i}_{00} + \chi^{i}_{10} q^{}_0+\chi^{i}_{20} q_0^2+ \chi^{i}_{01} p^{}_4 \nonumber \\
&+ \chi^{i}_{11} q^{}_0 p^{}_4 + \chi^{i}_{02} p_4^2,\label{eqn:eff_ex}
\end{align}
where $\chi^{}_i=\nu,~\phi$ and $\eta^{}_{\|,\perp}$.
The coefficients are presented in Table~\ref{tab:coeff_cr_exp}, and the results are plotted in Fig. \ref{exps}. Due to our quadratic approximations, these plots stop at $p^{}_4(\ell)=-.8$. However, the qualitative large variations are expected to continue towards the triple point.

\begin{table}[t]
 \centering
    \caption{Numerical values of coefficients entering Eq. (\ref{eqn:eff_ex})}.
    \label{tab:coeff_cr_exp}
     \setlength{\tabcolsep}{3.8pt}
    \begin{tabular}{ll|ll}
      \hline
      \hline
      Coeff. & Value & Coeff. & Value  \\
      \hline
 $\nu_{00}$& $0.70428(76)$ & $\eta^{\parallel}_{00}$ & $0.03762(85)$\\
 $\nu_{10}$& $0.731(10)$ & $\eta^{\parallel}_{10}$ & $0.1914(22)$\\
 $\nu_{01}$& $-0.0004(16)$ & $\eta^{\parallel}_{01}$ & $ 0.00038(22)$\\
 $\nu_{11}$& $-0.0009(69)$ & $\eta^{\parallel}_{11}$ & $-0.0081(32)$\\
 $\nu_{20}$& $2.369(61)$   & $\eta^{\parallel}_{20}$ & $0.630(18)$\\
 $\nu_{02}$& $-0.0610(24)$& $\eta^{\parallel}_{02}$ & $-0.00517(70)$\\
 $\phi_{00}$& $1.2614 (16)$   & $\eta^{\perp}_{00}$ & $0.03762(85)$\\
 $\phi_{10}$& $0.983(16)$   & $\eta^{\perp}_{10}$ & $0.1942(27)$\\
 $\phi_{01}$& $-0.2021(39)$  & $\eta^{\perp}_{01}$ & $0.00038(63)$\\
 $\phi_{11}$& $-0.203 (12)$  & $\eta^{\perp}_{11}$ & $-0.0033(87)$\\
 $\phi_{20}$& $3.314(83)$ & $\eta^{\perp}_{20}$ & $0.684(20)$\\
 $\phi_{02}$& $-0.1110 (34)$  & $\eta^{\perp}_{02}$ & $ 0.0106(14)$\\
    \hline
    \hline
    \end{tabular}
\end{table}

Note: For the small initial values of $p^{}_4$, all the exponents stay close to their isotropic FP values for a large range of $\ell$. This may explain the experimental and numerical observed results. However, as the correlation length increases the effective exponents deviate strongly from their isotropic values, until eventually the transition becomes first order at the triple point. Also, at the isotropic FP we must have  $\eta^{}_{\|}=\eta^{}_\perp$, and indeed $\eta^{\|}_{00}=\eta^\perp_{00}$.  Interestingly, $\eta^{\|}_{02}$ and $\eta^{\perp}_{02}$ have opposite signs. Since these terms dominate at large $\ell$, the deviation between the two $\eta$'s increases as the triple point is approached.

\section{ Conclusions}

Our accurate renormalization group calculations in the vicinity of the isotropic fixed point show that for a range of parameters, when $p^{}_4(0)<0$, the asymptotic multicritical point for the $n=2+1$ order parameters must cross over from the isotropic FP behavior to the triple point and the fluctuation-driven first order transition. Our calculated flow trajectories also allow us to calculate the effective critical exponents, which remain close to their isotropic values for a range of system sizes or correlation lengths, but then show large deviations as these length grow larger. Our results are qualitatively similar to those found for the cubic case,~\cite{AEK} and must also describe many other apparent bicritical points.~\cite{AE}

\bmhead{Acknowledgments}

We gratefully acknowledge A. Pikelner for the help with RG expansions. The work of A.K. was supported by Grant of the Russian
Science Foundation No 21-72-00108.

\bmhead{Data availability statements}

The coefficients of all RG expansions used in the paper are available at the "Ancillary files"  section on arXiv: \href{https://doi.org/10.48550/arXiv.2304.08265}{doi.org/10.48550/arXiv.2304.08265}.


\begin{thebibliography}{99}
\bibitem{SM} A full list of the numbers can be foud i the "ancillarey files" on ArXiv.230408265.
\bibitem{comvic}
 Instead of following the original RG calculations,~\cite{MEFAA} here we extended the field theoretical calculations of Refs..~\cite{vicari} and ~\cite{folk}, to sixth order in $\epsilon$. Using their definitions, we calculated the operator reccale factors for $S^{}_i$ ad $S^2_i$ as series in the $u^{}_i$'s, and derived from them the corrsponding resummed coefficients for the quadratic expansions of the various exponents near the isotropic FP.
\end{thebibliography}
\end{document}


\title[Article Title]{Effective exponents near bicritical points. Supplemental materials}


\author[1,2]{\fnm{Andrey} \sur{Kudlis}}\email{andrewkudlis@gmail.com}

\author*[3]{\fnm{Amnon} \sur{Aharony}}\email{aaharonyaa@gmail.com}

\author[3]{\fnm{Ora} \sur{Entin-Wohlman}}\email{orawohlman@gmail.com}


\affil[1]{\orgdiv{} \orgname{Abrikosov Center for Theoretical Physics, MIPT}, \orgaddress{\street{Kronverkskiy prospekt 49}, \city{Dolgoprudny}, \postcode{141701}, \state{Moscow Region}, \country{Russia}}}
\affil[2]{\orgdiv{} \orgname{ITMO University}, \orgaddress{\street{Kronverkskiy prospekt 49}, \city{Saint Petersburg}, \postcode{197101}, \state{State}, \country{Russia}}}

\affil*[3]{\orgdiv{School of Physics and Astronomy}, \orgname{Tel Aviv University}, \orgaddress{\street{Ramat Aviv}, \city{Tel Aviv}, \postcode{6997801}, \state{} \country{Israel}}}




\abstract{The phase diagram of a system with two order parameters, with ${\it n_1}$ and $n_2$ components, respectively, contains two phases, in which these order parameters are non-zero. Experimentally and numerically, these phases are often separated by a first-order ``flop" line, which ends at a bicritical point. For $n=n_1+n_2=3$ and $d=3$ dimensions (relevant e.g. to the uniaxial antiferromagnet in a uniform magnetic field), this bicritical point is found to exhibit a crossover from the isotropic $n$-component universal critical behavior to a fluctuation-driven first-order transition, asymptotically turning into a triple point. Using a novel expansion of the renormalization group recursion relations near the isotropic fixed point, combined with a resummation of the sixth-order diagrammatic expansions of the coefficients in this expansion, we show that the above crossover is slow, explaining the apparently observed second-order transition. However, the effective critical exponents near that transition, which are calculated here, vary strongly as the triple point is approached.

}

\keywords{bicritical point, renormalization group, effective exponents, fluctuation-driven first order transition, triple point}


\maketitle

Here we follow the notations of Refs.~\cite{vicari,folk}. Some normalization factors and signs in the definitions of quantities may differ, while the expressions for the observables coincide. As a starting point we expand the expressions (19), (22), and (32) in Ref.~\cite{folk} for the anomalous dimensions $\zeta_{\phi_{\parallel}^{}}$ ($\phi$ is the order parameter, which our paper denotes by $S$), $\zeta_{\phi_{\perp}^{}}$, and $\big[\zeta_{\phi_{}^2}\big]_{ij}$ and the  $\beta$-functions which we denote $\bar{\beta}^{}_{\parallel}$, $\bar{\beta}^{}_{\perp}$, and $\bar{\beta}^{}_{\times}$ up to six-loop contributions. This was done on the basis of the RG expansions for  general a scalar field theory, calculated in~\cite{bednyakov}, where the authors used available six-loop diagrams for the $O(n)$-symmetric model~\cite{kompanietz}. Using the anomalous dimensions, one can  derive expressions for effective exponents in terms of renormalized couplings:
\begin{align}\label{eqn:eta_in_terms_of_couplings}
&\eta^{}_{\parallel}=2\zeta^{}_{\phi^{}_{\parallel}}(u^{}_{\parallel},u^{}_{\perp},u^{}_{\times}),\\
&\eta^{}_{\perp}=2\zeta^{}_{\phi^{}_{\perp}}(u^{}_{\parallel},u^{}_{\perp},u^{}_{\times}).\label{eqn:eta_perp_in_terms_of_couplings}
\end{align}
As a self-check, we calculated six-loop expansions for the biconical fixed point (for $n^{}_{\parallel}=1$ and $n^{}_{\perp}=2$), which coincide with the known result up to two loops~\cite{folk}(up to a normalization factor 2). 
For the biconical fixed point, our expansions give
\begin{align}\label{eqn:biconical_f_p}
    u^{\ast,B}_{\parallel}&=0.20248 \epsilon+0.193457 \epsilon^2-0.864617 \epsilon^3-0.924397 \epsilon^4+2.30625 \epsilon^5+25.2141 \epsilon^6,\\
    u^{\ast,B}_{\perp}&=0.252845 \epsilon+0.15805 \epsilon^2+0.159863 \epsilon^3-0.17488 \epsilon^4+2.50834 \epsilon^5+7.76289 \epsilon^6,\\
    u^{\ast,B}_{\times}&=0.345294 \epsilon+0.14963 \epsilon^2+0.219463 \epsilon^3+1.51864 \epsilon^4-5.59288 \epsilon^5-22.4169 \epsilon^6,
\end{align}
where the number of digits is chosen arbitrarily, since all coefficients are known with absolute accuracy. Throughout this work, we fix  $n_{\parallel}^{}=1$ and $n_{\perp}^{}=2$. Combining~\eqref{eqn:eta_in_terms_of_couplings} and~\eqref{eqn:biconical_f_p}, the $\epsilon$ expansions for the critical exponents $\eta_{\parallel}^{}$($\eta_{\textup{bi},1}^{}$ in Ref.~\cite{vicari}) and $\eta_{\perp}$($\eta_{\textup{bi},2}$ in Ref.~\cite{vicari}) read as:
\begin{align}
    &\eta^{B}_{\parallel}=0.0208306 \epsilon^2+0.0182941 \epsilon^3-0.00777325 \epsilon^4+0.0210296 \epsilon^5-0.0595873 \epsilon^6+O(\epsilon^7),\\
   &\eta^{B}_{\perp}=0.0200806 \epsilon^2+0.0195184 \epsilon^3-0.0084802 \epsilon^4+0.0236398 \epsilon^5-0.0702837 \epsilon^6+O(\epsilon^7),
\end{align}
which coincide with (4.3) in Ref.~\cite{vicari}. We next diagoalize  the matrix $\big[\zeta^{}_{\phi_{}^2}\big]_{ij}^{}$,  obtaining the following eigenvalues:
\begin{align}\label{eqn:diag_an_dim}
&[\zeta^{\phi^2_{}}]^{}_{1} =
  \dfrac{1}{2} \left([\zeta_{\phi_{}^2}^{}]_{11}^{} + [\zeta^{}_{\phi^2_{}}]_{22}^{} - \sqrt{
     ([\zeta^{}_{\phi^2_{}}]_{11}^{})^2_{} + 4 [\zeta_{\phi^2_{}}^{}]_{12}^{}[\zeta_{\phi^2_{}}^{}]_{21}^{} - 2 [\zeta_{\phi^2_{}}^{}]_{11}^{} [\zeta_{\phi^2_{}}^{}]_{22}^{} + ([\zeta_{\phi^2_{}}^{}]_{22}^{})^2_{}}\right),\\
&[\zeta^{\phi^2_{}}]^{}_{2} =
  \dfrac{1}{2} \left([\zeta_{\phi^2_{}}^{}]_{11}^{} + [\zeta_{\phi^2_{}}^{}]_{22}^{} + \sqrt{
     ([\zeta_{\phi^2_{}}^{}]_{11}^{})^2_{} + 4 [\zeta^{}_{\phi^2_{}}]_{12}^{}[\zeta_{\phi^2_{}}^{}]_{21}^{} - 2 [\zeta_{\phi^2_{}}^{}]_{11}^{} [\zeta_{\phi^2_{}}^{}]_{22}^{} + ([\zeta_{\phi^2_{}}^{}]_{22}^{})^2_{}}\right),
\end{align}
all of which depend on the renormalized coupling constants. Based on these equations, one can obtain expressions for the effective critical exponents $\nu_{}$ and $\phi_{}$ as:
\begin{align}\label{eqn:exp_via_diag}
    \nu=\dfrac{1}{2+[\zeta^{\phi^2_{}}]^{}_{1}}, \quad \varphi=\dfrac{2+[\zeta^{\phi^2_{}}]^{}_{2}}{2+[\zeta^{\phi^2_{}}]^{}_{1}}.
\end{align}
At the  biconical fixed point, these expressions give
\begin{align}
    &\nu_{}^{B}=0.5 +0.111488 \epsilon+0.0667684 \epsilon^2-0.0061619 \epsilon^3+0.0779498 \epsilon^4-0.193367 \epsilon^5+0.698501 \epsilon^6+O(\epsilon^7),\\
    &\varphi_{}^{B}=1.+0.176147 \epsilon+0.0673541 \epsilon^2-0.0147318 \epsilon^3+0.0783198 \epsilon^4-0.190099 \epsilon^5+0.611865 \epsilon^6+O(\epsilon^7),
\end{align}
which coincide up to five loops with Eq (4.3) in Ref.~\cite{vicari}.

Starting from this point, we consider all RG expansions (which are known in terms of couplings $u_{\parallel}^{}$, $u_{\perp}^{}$, and $u_{\times}^{}$) in the vicinity of the isotropic Heisenberg fixed point. For this purpose, we need the $\epsilon$ expansions for this fixed point (within the chosen normalization):
\begin{align}
u_{\parallel}^{\ast,I}=u_{\perp}^{\ast,I}=u_{\times}^{\ast,I}=u^{\ast,I}_{}=0.2727 \epsilon+0.1555 \epsilon^2-0.08625 \epsilon^3+0.1615 \epsilon^4-0.4296 \epsilon^5+1.421 \epsilon^6+O(\epsilon^7).
\end{align}

In the vicinity of the isotropic fixed point the  exponents are expanded in the deviations $\delta u_{\parallel}^{}=(u_{\parallel}^{}-u_{}^{\ast,I})$, $\delta u_{\perp}^{}=(u_{\perp}^{}-u^{\ast,I})$, and $\delta u_{\times}^{}=(u_{\times}^{}-u_{}^{\ast,I})$, with coefficients which are derivatives of the $\bar{\beta}-$fuctions at the isotropic fixed point:
\begin{multline}
\eta_{\parallel}^{}(\delta u_{\parallel}^{},\delta u_{\perp}^{},\delta u_{\times}^{})=0.03762(85) -0.00043(43) \delta u_{\parallel}^{}-0.00048(60) \delta u_{\parallel}^2+0.1546(22) \delta u_{\perp}^{}-0.0061(16) \delta u_{\parallel}^{} \delta u_{\perp}^{}\\
+0.180(13) \delta u_{\perp}^2+0.03726(17) \delta u_{\times}^{}-0.00427(68) \delta u_{\parallel}^{} \delta u_{\times}^{}-0.006558(45) \delta u_{\perp}^{} \delta u_{\times}^{}+0.0570(32) \delta u_{\times}^2.
\end{multline}
\begin{multline}
\eta_{\perp}^{}(\delta u_{\parallel}^{},\delta u_{\perp}^{},\delta u_{\times}^{})=0.03762(85) + 0.1181(22) \delta u_{\parallel}^{} + 0.178(13) \delta u_{\parallel}^2 - 0.0012(11) \delta u_{\perp}^{} -  0.0123(32) \delta u_{\parallel}^{} \delta u_{\perp}^{}\\
- 0.0016(22) \delta u_{\perp}^2 + 0.0.0773(11) \delta u_{\times}^{} -  0.00679(81) \delta u_{\parallel}^{} \delta u_{\times}^{} - 0.0114(18) \delta u_{\perp}^{} \delta u_{\times}^{} + 0.1227(94) \delta u_{\times}^2,
\end{multline}
\begin{multline}
\nu(\delta u_{\parallel}^{},\delta u_{\perp}^{},\delta u_{\times}^{})=0.70428(76) +0.1464(32) \delta u_{\parallel}^{}+0.237(13) \delta u_{\parallel}^2+0.3885(92) \delta u_{\perp}^{}-0.422(14) \delta u_{\parallel}^{} \delta u_{\perp}^{}\\+0.603(36) \delta u_{\perp}^2
+0.1960(34) \delta u_{\times}^{}+0.2647(67) \delta u_{\parallel}^{} \delta u_{\times}^{}+0.0759(44) \delta u_{\perp}^{} \delta u_{\times}^{} +0.0443(54) \delta u_{\times}^2,
\end{multline}
\begin{multline}
\varphi(\delta u_{\parallel}^{},\delta u_{\perp}^{},\delta u_{\times}^{})=1.2614(16) -0.1085(52) \delta u_{\parallel}^{}+0.646(29) \delta u_{\parallel}^2+0.460(11) \delta u_{\perp}^{}-1.842(60) \delta u_{\parallel}^{} \delta u_{\perp}^{}\\+1.478(76) \delta u_{\perp}^2
+0.632(11) \delta u_{\times}^{}+0.729(17) \delta u_{\parallel}^{} \delta u_{\times}^{}+0.108(13) \delta u_{\perp}^{} \delta u_{\times}^{}+0.089(15) \delta u_{\times}^2.
\end{multline}

Let us show in detail how we obtained these expressions using the example of the exponent $\eta_{\perp}^{}(\delta u_{\parallel}^{},\delta u_{\perp}^{},\delta u_{\times}^{})$. First, we derive similar expansions for $\zeta_{\phi_{\parallel}^{}}^{}$, $\zeta_{\phi_{\perp}^{}}^{}$, and $[\zeta_{\phi_{}^2}]_{ij}^{}$. Then we obtain series for effective exponents by means of~\eqref{eqn:eta_in_terms_of_couplings},~\eqref{eqn:eta_perp_in_terms_of_couplings},~\eqref{eqn:diag_an_dim} and~\eqref{eqn:exp_via_diag}. Thus, as example, we should consider $\zeta_{\phi_{\perp}^{}}^{}(\delta u_{\parallel}^{},\delta u_{\perp}^{},\delta u_{\times}^{})$. As explaied in the main text, we limit ourselves to the quadratic order in the $\delta u$'s. In this case, the expansion reads:
\begin{multline}
\zeta_{\phi_{\perp}^{}}^{}(\delta u_{\parallel}^{},\delta u_{\perp}^{},\delta u_{\times}^{})=\zeta_{\phi_{\perp}^{}}^{000}+\zeta_{\phi_{\perp}^{}}^{100} \delta u_{\parallel}+\zeta_{\phi_{\perp}^{}}^{010}\delta u_{\perp}+\zeta_{\phi_{\perp}^{}}^{001} \delta u_{\times}+ \zeta_{\phi_{\perp}^{}}^{200}/2 \delta u_{\parallel}^2+\zeta_{\phi_{\perp}^{}}^{110}\delta u_{\parallel}\delta u_{\perp}\\
+ \zeta_{\phi_{\perp}^{}}^{020}/2 \delta u_{\perp}^2+\zeta_{\phi_{\perp}^{}}^{011} \delta u_{\perp} \delta u_{\times}+\zeta_{\phi_{\perp}^{}}^{002}/2 \delta u_{\times}^2+\zeta_{\phi_{\perp}^{}}^{101} \delta u_{\parallel} \delta u_{\times},
\end{multline}
where the coefficients were obtained from resummation (see Ref.~\cite{kompanietz} for details of the resummation procedure) of $\epsilon$ expansions in the vicinity of the isotropic Heisenberg fixed point:
\allowdisplaybreaks
\begin{align}
    \zeta_{\phi_{\perp}^{}}^{000}&=\text{Resum}\Bigg[\left.\zeta_{\phi_{\perp}^{}}(u_{\parallel},u_{\perp},u_{\times})\right\vert_{u_{\parallel}\rightarrow u_{\parallel}^{\ast,I},u_{\perp}\rightarrow u_{\perp}^{\ast,I},u_{\times}\rightarrow u_{\times}^{\ast,I}} \Bigg]\nonumber\\
    &=\text{Resum}\Bigg[0.0103306  \epsilon^2+0.00919934  \epsilon^3-0.00372474  \epsilon^4+0.0101914  \epsilon^5-0.0285121  \epsilon^6  \Bigg],\\
    \zeta_{\phi_{\perp}^{}}^{100}&=\text{Resum}\Bigg[\left.\dfrac{\partial\zeta_{\phi_{\perp}^{}}(u_{\parallel},u_{\perp},u_{\times})}{\partial u_{\parallel} }\right\vert_{u_{\parallel}\rightarrow u_{\parallel}^{\ast,I},u_{\perp}\rightarrow u_{\perp}^{\ast,I},u_{\times}\rightarrow u_{\times}^{\ast,I}} \Bigg]\nonumber\\
    &=\text{Resum}\Bigg[0.0454545 \epsilon+0.00990796 \epsilon^2+0.000357379 \epsilon^3+0.0173895 \epsilon^4-0.059075 \epsilon^5\Bigg],\\
    \zeta_{\phi_{\perp}^{}}^{010}&=\text{Resum}\Bigg[\left.\dfrac{\partial\zeta_{\phi_{\perp}^{}}(u_{\parallel},u_{\perp},u_{\times})}{\partial u_{\times} }\right\vert_{u_{\parallel}\rightarrow u_{\parallel}^{\ast,I},u_{\perp}\rightarrow u_{\perp}^{\ast,I},u_{\times}\rightarrow u_{\times}^{\ast,I}} \Bigg]\nonumber\\
    &=\text{Resum}\Bigg[-0.00137741 \epsilon^2-0.000172651 \epsilon^3-0.0064207 \epsilon^4+0.00745328 \epsilon^5\Bigg],\\
     \zeta_{\phi_{\perp}^{}}^{001}&=\text{Resum}\Bigg[\left.\dfrac{\partial\zeta_{\phi_{\perp}^{}}(u_{\parallel},u_{\perp},u_{\times})}{\partial u_{\times} }\right\vert_{u_{\parallel}\rightarrow u_{\parallel}^{\ast,I},u_{\perp}\rightarrow u_{\perp}^{\ast,I},u_{\times}\rightarrow u_{\times}^{\ast,I}} \Bigg]\nonumber\\
    &=\text{Resum}\Bigg[0.030303 \epsilon+0.00626096 \epsilon^2+0.00019509 \epsilon^3+0.00998786 \epsilon^4-0.03752 \epsilon^5\Bigg],\\
    \zeta_{\phi_{\perp}^{}}^{200}&=\text{Resum}\Bigg[\left.\dfrac{\partial^2\zeta_{\phi_{\perp}^{}}(u_{\parallel}^{},u_{\perp}^{},u_{\times}^{})}{\partial u_{\parallel}^2 }\right\vert_{u_{\parallel}^{}\rightarrow u_{\parallel}^{\ast,I},u_{\perp}^{}\rightarrow u_{\perp}^{\ast,I},u_{\times}^{}\rightarrow u_{\times}^{\ast,I}} \Bigg]\nonumber\\
    &=\text{Resum}\Bigg[0.166667 -0.102273 \epsilon+0.255729 \epsilon^2-0.462762 \epsilon^3+1.12195 \epsilon^4\Bigg],\\
    \zeta_{\phi_{\perp}^{}}^{110}&=\text{Resum}\Bigg[\left.\dfrac{\partial^2\zeta_{\phi_{\perp}^{}}(u_{\parallel}^{},u_{\perp}^{},u_{\times}^{})}{\partial u_{\parallel}^{}\partial u_{\times}^{} }\right\vert_{u_{\parallel}^{}\rightarrow u_{\parallel}^{\ast,I},u_{\perp}^{}\rightarrow u_{\perp}^{\ast,I},u_{\times}\rightarrow u_{\times}^{\ast,I}} \Bigg]\nonumber\\
    &=\text{Resum}\Bigg[-0.00344353 \epsilon^2-0.00684332 \epsilon^3+0.0216822 \epsilon^4\Bigg],\\
    \zeta_{\phi_{\perp}^{}}^{011}&=\text{Resum}\Bigg[\left.\dfrac{\partial^2\zeta_{\phi_{\perp}^{}}^{}(u_{\parallel}^{},u_{\perp}^{},u_{\times}^{})}{\partial u_{\perp}^{}\partial u_{\times}^{} }\right\vert_{u_{\parallel}^{}\rightarrow u_{\parallel}^{\ast,I},u_{\perp}^{}\rightarrow u_{\perp}^{\ast,I},u_{\times}^{}\rightarrow u_{\times}^{\ast,I}} \Bigg]\nonumber\\
    &=\text{Resum}\Bigg[-0.010101 \epsilon+0.0162785 \epsilon^2-0.0772736 \epsilon^3+0.187738 \epsilon^4\Bigg],\\
    \zeta_{\phi_{\perp}^{}}^{101}&=\text{Resum}\Bigg[\left.\dfrac{\partial^2\zeta_{\phi_{\perp}^{}}(u_{\parallel}^{},u_{\perp}^{},u_{\times}^{})}{\partial u_{\parallel}^{}\partial u_{\times}^{} }\right\vert_{u_{\parallel}^{}\rightarrow u_{\parallel}^{\ast,I},u_{\perp}^{}\rightarrow u_{\perp}^{\ast,I},u_{\times}^{}\rightarrow u_{\times}^{\ast,I}} \Bigg]\nonumber\\
    &=\text{Resum}\Bigg[-0.0151515 \epsilon+0.0437015 \epsilon^2-0.119204 \epsilon^3+0.316655 \epsilon^4\Bigg],\\
    \zeta_{\phi_{\perp}^{}}^{020}&=\text{Resum}\Bigg[\left.\dfrac{\partial^2\zeta_{\phi_{\perp}^{}}(u_{\parallel}^{},u_{\perp}^{},u_{\times}^{})}{\partial u_{\perp}^2 }\right\vert_{u_{\parallel}^{}\rightarrow u_{\parallel}^{\ast,I},u_{\perp}^{}\rightarrow u_{\perp}^{\ast,I},u_{\times}^{}\rightarrow u_{\times}^{\ast,I}} \Bigg]\nonumber\\
    &=\text{Resum}\Bigg[-0.00321396 \epsilon^2-0.0306099 \epsilon^3+0.0911289 \epsilon^4\Bigg],\\
    \zeta_{\phi_{\perp}^{}}^{002}&=\text{Resum}\Bigg[\left.\dfrac{\partial^2\zeta_{\phi_{\perp}^{}}(u_{\parallel}^{},u_{\perp}^{},u_{\times}^{})}{\partial u_{\times}^2 }\right\vert_{u_{\parallel}^{}\rightarrow u_{\parallel}^{\ast,I},u_{\perp}^{}\rightarrow u_{\perp}^{\ast,I},u_{\times}^{}\rightarrow u_{\times}^{\ast,I}} \Bigg]\nonumber\\
    &=\text{Resum}\Bigg[0.111111 -0.0555556 \epsilon+0.13975 \epsilon^2-0.224743 \epsilon^3+0.544271 \epsilon^4\Bigg],
\end{align}
where all the $\epsilon$ expansion coefficients are known with absolute accuracy. In the same way one can obtain the other expansions,  for $\zeta_{\phi_{\parallel}^{}}^{}(\delta u_{\parallel}^{},~\delta u_{\perp}^{},\delta u_{\times}^{})$, $[\zeta_{\phi_{}^2}^{}]_{ij}^{}(\delta u_{\parallel}^{},\delta u_{\perp}^{},\delta u_{\times}^{})$  and their coefficients are extracted from resummations of corresponding $\epsilon$ series. These are presented in Table~\ref{tab:coef_ef_exp_u}.
\begin{table}[h]
 \centering
    \caption{Numerical values of expansion coefficients of anomalous dimensions in the vicinity of the isotropic Heisenberg fixed point.} %
    \label{tab:coef_ef_exp_u}
     \setlength{\tabcolsep}{10.0pt}
    \begin{tabular}{l|l|l|l}
      \hline
      \hline
$[\zeta_{\phi^2_{}}]_{11}^{000}=-0.33322(96)$&$ [\zeta_{\phi^2_{}}]_{11}^{020}=0.00188(58)$&$ [\zeta_{\phi_{}^2}]_{12}^{000}=-0.12383(98)$&$ [\zeta_{\phi_{}^2}]_{12}^{020}=0.0494(33)$\\$[\zeta_{\phi_{}^2}]_{21}^{000}=-0.2477(20)$&$ [\zeta_{\phi_{}^2}]_{21}^{020}=0.0589(59)$&$ [\zeta_{\phi_{}^2}]_{22}^{000}=-0.4559(18)$&$ [\zeta_{\phi_{}^2}]_{22}^{020}=-0.07(14)$\\$\zeta_{\phi_{\perp}^{}}^{000}=0.01881(42)$&$ \zeta_{\phi_{\perp}^{}}^{020}=-0.0016(22)$&$ \zeta_{\phi_{\parallel}^{}}^{000}=0.01881(42)$&$ \zeta_{\phi_{\parallel}^{}}^{020}=0.180(13)$\\$[\zeta_{\phi_{}^2}]_{11}^{001}=0.0075(25)$&$ [\zeta_{\phi_{}^2}]_{11}^{100}=-0.830(17)$&$ [\zeta_{\phi_{}^2}]_{12}^{001}=-0.2964(70)$&$ [\zeta_{\phi_{}^2}]_{12}^{100}=-0.0177(15)$\\$[\zeta_{\phi_{}^2}]_{21}^{001}=-0.592(14)$&$ [\zeta_{\phi_{}^2}]_{21}^{100}=-0.0336(31)$&$ [\zeta_{\phi_{}^2}]_{22}^{001}=-0.0036(18)$&$ [\zeta_{\phi_{}^2}]_{22}^{100}=0.0087(21)$\\$\zeta_{\phi_{\perp}^{}}^{001}=0.03865(54)$&$ \zeta_{\phi_{\perp}^{}}^{100}=0.0590(11)$&$ \zeta_{\phi_{\parallel}^{}}^{001}=0.018632(86)$&$ \zeta_{\phi_{\parallel}^{}}^{100}=-0.00022(21)$\\$[\zeta_{\phi_{}^2}]_{11}^{002}=-0.106(17)$&$ [\zeta_{\phi_{}^2}]_{11}^{101}=0.0189(96)$&$ [\zeta_{\phi_{}^2}]_{12}^{002}=0.150(14)$&$ [\zeta_{\phi_{}^2}]_{12}^{101}=-0.0766(19)$\\$[\zeta_{\phi_{}^2}]_{21}^{002}=0.308(31)$&$ [\zeta_{\phi_{}^2}]_{21}^{101}=-0.1769(35)$&$ [\zeta_{\phi_{}^2}]_{22}^{002}=-0.046(17)$&$ [\zeta_{\phi_{}^2}]_{22}^{101}=0.0368(20)$\\$\zeta_{\phi_{\perp}^{}}^{002}=0.1227(94)$&$ \zeta_{\phi_{\perp}^{}}^{101}=-0.00339(40)$&$ \zeta_{\phi_{\parallel}^{}}^{002}=0.0570(32)$&$ \zeta_{\phi_{\parallel}^{}}^{101}=-0.00214(34)$\\$[\zeta_{\phi_{}^2}]_{11}^{010}=0.0249(62)$&$ [\zeta_{\phi_{}^2}]_{11}^{110}=-0.0035(31)$&$ [\zeta_{\phi_{}^2}]_{12}^{010}=-0.0226(23)$&$ [\zeta_{\phi_{}^2}]_{12}^{110}=0.00143(33)$\\$[\zeta_{\phi_{}^2}]_{21}^{010}=-0.0476(46)$&$ [\zeta_{\phi_{}^2}]_{21}^{110}=0.00294(62)$&$ [\zeta_{\phi_{}^2}]_{22}^{010}=-1.143(27)$&$ [\zeta_{\phi_{}^2}]_{22}^{110}=-0.0017(14)$\\$\zeta_{\phi_{\perp}^{}}^{010}=-0.00058(57)$&$ \zeta_{\phi_{\perp}^{}}^{110}=-0.0062(16)$&$ \zeta_{\phi_{\parallel}^{}}^{010}=0.0773(11)$&$ \zeta_{\phi_{\parallel}^{}}^{110}=-0.00305(80)$\\$[\zeta_{\phi_{}^2}]_{11}^{011}=0.1093(73)$&$ [\zeta_{\phi_{}^2}]_{11}^{200}=0.010(91)$&$ [\zeta_{\phi_{}^2}]_{12}^{011}=-0.1150(28)$&$ [\zeta_{\phi_{}^2}]_{12}^{200}=0.0221(24)$\\$[\zeta_{\phi_{}^2}]_{21}^{011}=-0.2092(37)$&$ [\zeta_{\phi_{}^2}]_{21}^{200}=0.0803(60)$&$ [\zeta_{\phi_{}^2}]_{22}^{011}=0.0074(46)$&$ [\zeta_{\phi_{}^2}]_{22}^{200}=0.00081(23)$\\$\zeta_{\phi_{\perp}^{}}^{011}=-0.00568(89)$&$ \zeta_{\phi_{\perp}^{}}^{200}=0.178(13)$&$ \zeta_{\phi_{\parallel}^{}}^{011}=-0.003279(23)$&$ \zeta_{\phi_{\parallel}^{}}^{200}=-0.00048(60)$\\
    \hline
    \hline
    \end{tabular}
\end{table}
In order to track the behavior of the effective critical exponents, we need to write the recursion relations for the couplings. To this purpose, the six-loop expansions for the $\bar{\beta}$-functions were obtained and re-expanded in the vicinity of the isotropic Heisenberg fixed point. The procedure is the same as described above. The obtained expansions read as:
\begin{multline}
-\beta_{\parallel}^{}=L^{\parallel}_{100}\delta u_{\parallel}^{}+L^{\parallel}_{010}\delta u_{\perp}^{}+L^{\parallel}_{001}\delta u_{\times}^{} +L^{\parallel}_{200}/2\delta u_{\parallel}^2+L^{\parallel}_{020}/2\delta u_{\perp}^2+L^{\parallel}_{002}/2\delta u_{\times}^2\\
+L^{\parallel}_{110}\delta u_{\parallel}^{}\delta u_{\perp}^{}+L^{\parallel}_{101}\delta u_{\parallel}^{}\delta u_{\times}^{}+L^{\parallel}_{011}\delta u_{\perp}^{}\delta u_{\times}^{},
\end{multline}
\begin{multline}
-\beta_{\perp}^{}=L^{\perp}_{100}\delta u_{\parallel}^{}+L^{\perp}_{010}\delta u_{\perp}^{}+L^{\perp}_{001}\delta u_{\times}^{}
+L^{\perp}_{200}/2\delta u_{\parallel}^2+L^{\perp}_{020}/2\delta u_{\perp}^2+L^{\perp}_{002}/2\delta u_{\times}^2\\
+L^{\perp}_{110}\delta u_{\parallel}^{}\delta u_{\perp}^{}+L^{\perp}_{101}\delta u_{\parallel}^{}\delta u_{\times}^{}+L^{\perp}_{011}\delta u_{\perp}^{}\delta u_{\times}^{},
\end{multline}
\begin{multline}
-\beta_{\times}^{}=L^{\times}_{100}\delta u_{\parallel}^{}+L^{\times}_{010}\delta u_{\perp}^{}+L^{\times}_{001}\delta u_{\times}^{}
+L^{\times}_{200}/2\delta u_{\parallel}^2+L^{\times}_{020}/2\delta u_{\perp}^2+L^{\times}_{002}/2\delta u_{\times}^2\\
L^{\times}_{110}\delta u_{\parallel}^{}\delta u_{\perp}^{}+L^{\times}_{101}\delta u_{\parallel}^{}\delta u_{\times}^{}+L^{\times}_{011}\delta u_{\perp}^{}\delta u_{\times}^{}.
\end{multline}
The numerical values of the coefficients are listed in Table~\ref{tab:coef_beta}.
\begin{table}[h]
 \centering
    \caption{Numerical values of expansion coefficients of anomalous dimensions in the vicinity of the isotropic Heisenberg fixed point.} %
    \label{tab:coef_beta}
     \setlength{\tabcolsep}{38.4pt}
    \begin{tabular}{l|l|l}
      \hline
      \hline
      $L^{\parallel}_{001}=0.3205(11)$&$ L^{\parallel}_{002}=0.852(54)$&$ L^{\parallel}_{010}=0.0020(13)$\\$L^{\parallel}_{011}=0.011(33)$&$ L^{\parallel}_{020}=-0.0105(24)$&$ L^{\parallel}_{100}=0.4761(10)$\\$L^{\parallel}_{101}=-0.249(97)$&$ L^{\parallel}_{110}=-0.074(14)$&$ L^{\parallel}_{200}=3.118(76)$\\$L^{\perp}_{001}=0.1592(18)$&$ L^{\perp}_{002}=0.427(23)$&$ L^{\perp}_{010}=0.6346(15)$\\$L^{\perp}_{011}=-0.139(38)$&$ L^{\perp}_{020}=3.268(37)$&$ L^{\perp}_{100}=0.00077(47)$\\$L^{\perp}_{101}=0.004(12)$&$ L^{\perp}_{110}=-0.0255(48)$&$ L^{\perp}_{200}=-0.0050(12)$\\$L^{\times}_{001}=0.229(10)$&$ L^{\times}_{002}=1.306(32)$&$ L^{\times}_{010}=0.3183(36)$\\$L^{\times}_{011}=0.605(66)$&$ L^{\times}_{020}=0.040(13)$&$ L^{\times}_{100}=0.24039(79)$\\$L^{\times}_{101}=0.469(26)$&$ L^{\times}_{110}=-0.0044(66)$&$ L^{\times}_{200}=0.026(13)$\\
    \hline
    \hline
    \end{tabular}
\end{table}

Having obtained the coefficients $L^i_{abc}$, Eq. (9) in the main text allows us to find the recursion relations for the $p^{}_i$'s, which are icluded i the main text for the special case $p_2^{}\equiv 0$.

\begin{thebibliography}{99}

\bibitem{vicari} P. Calabrese, A. Pelissetto, and E. Vicari, {\it Multicritical phenomena in $O(n^{}_1)\bigoplus O(n^{}_2)$-symmetric theories}, 
    Phys. Rev. B {\bf 67}, 054505 (2003).

\bibitem{folk} R. Folk, Yu. Holovatch, and G. Moser, {\it Field theory of bicritical and tetracritical points. I. Statics}, 
    Phys. Rev. E {\bf 78}, 041124 (2008). 

\bibitem{bednyakov} A. Bednyakov and A. Pikelner, {\it Six-loop beta functions in general scalar theory}, Journal of High Energy
Physics, {\bf 2021}, 233 (2021).

\bibitem{kompanietz} Mikhail V. Kompaniets and Erik Panzer, {\it Minimally subtracted six-loop renormalization of O(n)-symmetric
$\phi^4$ theory and critical exponents}, Phys. Rev. D {\bf 96}, 036016 (2017).